\RequirePackage{fix-cm}
\documentclass[twocolumn]{svjour3}          
\smartqed  
\usepackage{graphicx}
\usepackage{cite}
\usepackage{amsmath,amssymb}
\usepackage[dvipsnames]{xcolor}
\usepackage{hyperref}
\usepackage[dvipsnames]{xcolor}
\usepackage[markup=nocolor]{changes}
\usepackage{soul}
\usepackage{xcolor}

\begin{document}
\title{Regimes of cosmic-ray diffusion in Galactic turbulence}

\author{P.~Reichherzer$^{1,2,3}$ \and L.~Merten$^{4}$ \and J.~Dörner$^{1,2}$ \and J.~Becker Tjus$^{1,2}$ \and  M.J.~Pueschel$^{5,6}$ \and E.G.~Zweibel$^{7,8}$ \\\\
$^1$ Theoretical Physics IV: Plasma-Astroparticle Physics, Faculty for Physics \& Astronomy, Ruhr-Universit\"at Bochum, D-44780 Bochum, Germany\\
$^2$ Ruhr Astroparticle And Plasma Physics Center (RAPP Center)\\
$^3$ IRFU, CEA, Université Paris-Saclay, F-91191 Gif-sur-Yvette, France\\
$^4$ Institute for Astro- \& Particle Physics, University of Innsbruck, 6020 Innsbruck, Austria\\
$^5$ Dutch Institute for Fundamental Energy Research, 5612 AJ Eindhoven, The Netherlands,\\
$^6$ Eindhoven University of Technology, 5600 MB Eindhoven, The Netherlands,\\
$^7$ Department of Physics, University of Wisconsin-Madison, Madison, WI 53706, U.S.A.\\
$^8$ Department of Astronomy, University of Wisconsin-Madison, Madison, WI 53706, U.S.A.}


\institute{Patrick Reichherzer \at
              \email{patrick.reichherzer@ruhr-uni-bochum.de}           
}

\date{Accepted: 22 November 2021 / Published: 11 December 2021 / \href{https://link.springer.com/article/10.1007/s42452-021-04891-z}{\textit{SN Appl. Sci.} \textbf{4}, 15 (2022)}}

\maketitle
\begin{abstract}
Cosmic-ray transport in astrophysical environments is often dominated by the diffusion of particles in a magnetic field composed of both a turbulent and a mean component. This process, which is two-fold turbulent mixing in that the particle motion is stochastic with respect to the field lines, needs to be understood in order to properly model cosmic-ray signatures. One of the most important aspects in the modeling of cosmic-ray diffusion is that fully resonant scattering, the most effective such process, is only possible if the wave spectrum covers the entire range of propagation angles.
By taking the wave spectrum boundaries into account, we quantify cosmic-ray diffusion parallel and perpendicular to the guide field direction at turbulence levels above 5\% of the total magnetic field. 
We apply our results of the parallel and perpendicular diffusion coefficient to the Milky Way. We show that simple purely diffusive transport is in conflict with observations of the inner Galaxy, but that just by taking a Galactic wind into account, data can be matched in the central 5 kpc zone. Further comparison shows that the outer Galaxy at $>5$~kpc, on the other hand, should be dominated by perpendicular diffusion, likely changing to parallel diffusion at the outermost radii of the Milky Way.
\end{abstract}

\section{Introduction}
\label{intro}
The origin of cosmic rays has been the subject of much research since the first detection of cosmic rays in 1912, see e.g.\,\cite{BeckerTjus2020} for a review. 
Baade \& Zwicky's \cite{baade_zwicky1934} hypothesis that supernovae are the most likely energy source of cosmic rays has been strengthened, and significant advances have been made in understanding particle acceleration in supernova remnant environments \cite{BeckerTjus2020}. Equally important
is a thorough understanding of cosmic-ray transport and interactions. In this paper, the focus lies on the understanding of the cosmic-ray diffusion process, which determines the evolution of the density of cosmic rays in a spatially limited environment like the Milky Way. 
In general, turbulent mixing of collisionless charged particles in magnetized plasma offers interesting challenges not encountered in hydrodynamic turbulent mixing, and thus is highly appropriate to this topical collection. Due to the high electrical conductivity of plasma, the magnetic field itself is mixed by turbulence, and the field lines may become stochastic \cite{Olivier2019}. To the extent that cosmic rays, being electrically charged particles, follow the field lines, they are spatially mixed the same way the field lines are. In addition, because charged particle orbits are helical, with an energy-dependent size, and because cosmic rays are scattered by orbit scale magnetic fluctuations, they both cross field lines and reverse the direction of motion along them. Moreover, the problem is nonlinear in the sense that the cosmic rays modify their environment by heating, pressurizing, and transferring momentum to the thermal background gas, and by feeding back on the rate of scattering itself. Because the time scales for all these processes are similar to the time scales on which the entire astrophysical system evolves, the problem is inherently nonequilibrium. Developing a full description of the resulting spatial mixing process is one of the most challenging problems in plasma physics, and is necessary not only to understand the transport of cosmic rays, but also particle confinement in laboratory plasma devices, and thermal conductivity and viscosity in both lab and natural plasmas.

Diffusion is usually incorporated through the Parker transport equation
\begin{align}
    \frac{\partial n}{\partial t} + \vec{u}\cdot\nabla n &= \nabla\cdot(\hat{\kappa}\nabla n) + \frac{1}{p^2}\frac{\partial}{\partial p}\left(p^2\kappa_{pp}\frac{\partial n}{\partial p}\right)  \notag \\
    &+ \frac{p}{3}(\nabla\cdot \vec{u})\frac{\partial n}{\partial p} + S\quad . \label{eq:ParkerTransport}
\end{align}
Here, $\vec{u}$ is an advection speed, $p$ is the momentum, $n$ is the particle distribution, $\hat{\kappa}$ is the spatial diffusion tensor, and $\kappa_{pp}$ is the momentum diffusion scalar. In this equation, we have assumed isotropy in momentum space. While in the Parker transport equation, sources and sinks of cosmic rays are included within the term $S$, all losses due to interactions, such as synchrotron radiation or spallation of nuclei on the ambient interstellar medium, are neglected. 

\begin{figure*}
\centering
\includegraphics[width=0.9\linewidth]{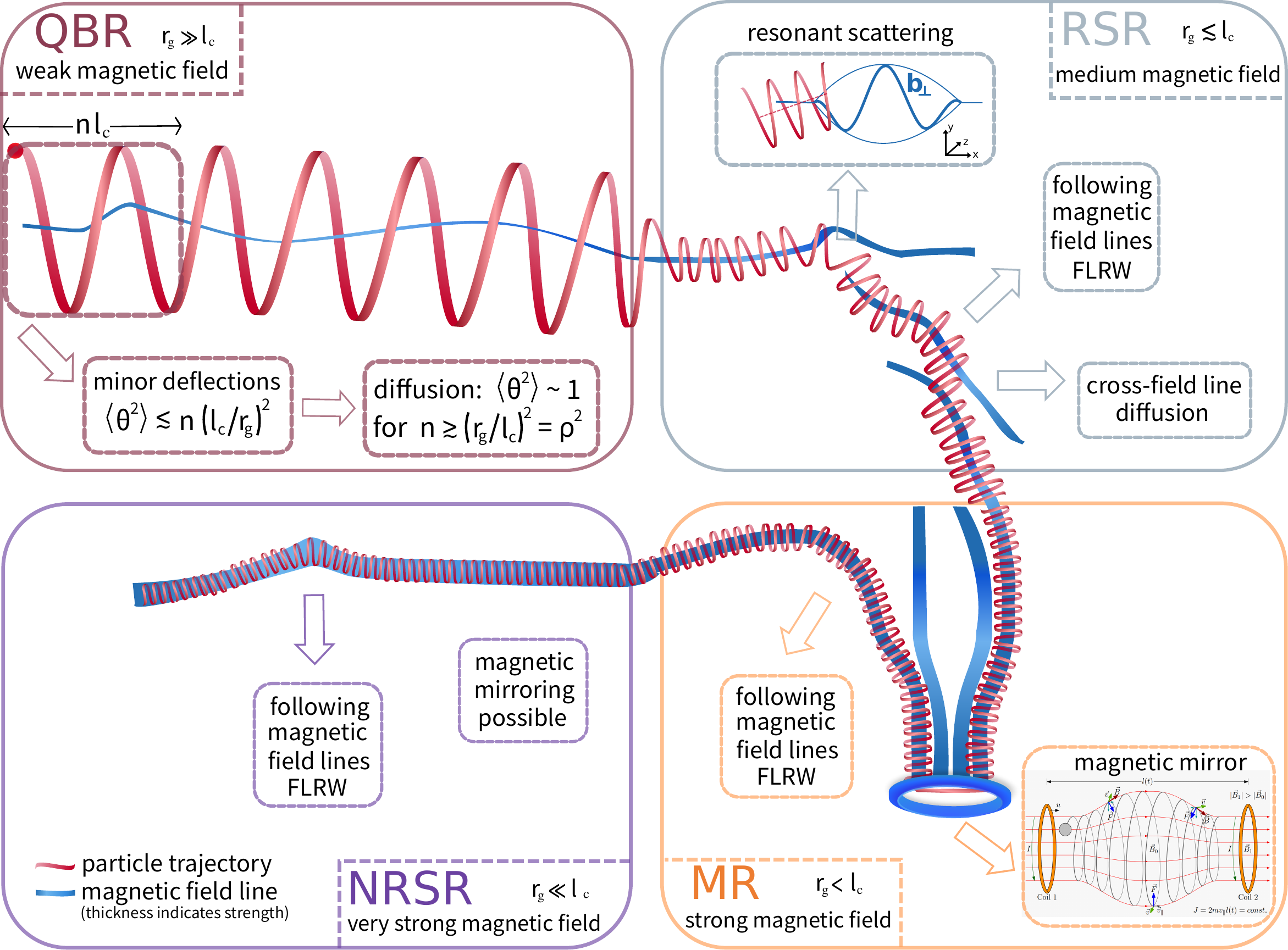}
\caption{
Schematic picture of a charged particle traveling through different magnetic field strengths. The magnetic field strength increases starting from the upper left panel and proceeding clockwise, resulting in an accompanying decrease of the particle gyroradius. The ratio of gyroradius and correlation length determines the predominant diffusion regime of the particle. The possible regimes are the quasi-ballistic regime (QBR), the resonant-scattering regime (RSR), the mirror regime (MR), and the non-resonant-scatter regime (NRSR) \cite{Reichherzer2020}. Equivalently, the different regimes can also be represented schematically by changing particle energy between the boxes at constant magnetic field strengths. In each regime, different processes dominate diffusion, such as FLRW \cite{Minnie2009} and the transport across magnetic field lines \cite{Desiati2014}.} The condition for diffusion in QBR follows from \cite{Subedi2017} and is generalized here for the inclusion of background fields. $n$ stands for the number of boxes with side length $l_c$. The box and thus the scale that would be necessary to allow diffusion exceeds the width of the region shown in the panel, as $n l_\mathrm{c} \gtrsim r_\mathrm{g}/l_\mathrm{c} \gg r_\mathrm{g}$. The mean-free path is approximately given by $nl_\mathrm{c}$, for the smallest $n$ which satisfies the condition for diffusion.\label{fig:fig1}
\end{figure*}

Cosmic-ray diffusion is believed to be the dominant process for the transport of cosmic rays in many astrophysical environments \cite{galprop,dragon,picard, Mertsch2020}. However, the components of the spatial diffusion tensor are also the most elusive parameters in Eq.\ (\ref{eq:ParkerTransport}), because they depend on a variety of physical processes: Depending on the ratio of the gyroradius to the correlation length $l_c$ of the turbulence, certain effects dominate in the diffusion process, leading to five different transport regimes, see \cite{Reichherzer2020}. In Fig.~\ref{fig:fig1}, this is shown schematically based on the example of a charged particle that moves through different magnetic field configurations. Along the trajectory, the magnetic field strength between the boxes changes, so that four transport regimes (the 5th regime is a transition between two other regimes) are illustrated. The diagram illustrates how the motion of the particle in the quasi-ballistic regime (QBR) is largely unaffected by the specifics of the field lines as a result of the large gyroradius. The decrease of the gyroradius results in a stronger influence of the magnetic field lines on the diffusion properties of the charged particles, because they are tied closely to the lines, but also because the effects of the cross-field motion become more important up to a certain level. The diffusivity of the field line itself adds to the diffusivity of the following particles due to random walk. Overall, resonant scattering dominates the diffusion process, especially in the RSR. Resonant scattering of particles with a pitch angle $\Theta_0$ at fluctuations of wavelength $l$ is present only when the resonant criterion is fulfilled
\begin{align}\label{resonantScattering_0}
|\mu| = \frac{l}{2\pi r_{\mathrm{g}}},
\end{align}
where $\mu = \cos\Theta_0$ is defined as the cosine of the pitch angle. To what extent mirror effects towards lower energies compensate the absence of resonant scattering and lead to diffusion remains a subject of active research \cite{Cesarsky1973, felice2001, lange2013, seta2018}.

Within a magnetic field $\vec{B}_\mathrm{tot}$, which is composed of a coherent component along the x-direction $\vec{B}=B\vec{e}_x$ and a turbulent component $\vec{b}$, the diffusion tensor can be written in matrix form \cite{Giacalone1999, Snodin2015, sigl2017, BeckerTjus2020}
\begin{align}
\hat{\kappa} = 
\begin{pmatrix}
 \kappa_\parallel & 0 & 0 \\
 0 & \kappa_\perp & \kappa_A \\
 0 & -\kappa_A & \kappa_\perp
\end{pmatrix} \quad .
\end{align}
Transport along the coherent magnetic field is described by the parallel diffusion coefficient $\kappa_\parallel \equiv \kappa_{xx}$ and is complemented by the perpendicular $\kappa_\perp \equiv \kappa_{yy} = \kappa_{zz}$.\footnote{In general, the two perpendicular directions can be different, $\kappa_{yy}\neq\kappa_{zz}$. However, within the Galactic magnetic field, they are usually degenerate, $\kappa_{yy}=\kappa_{zz}=\kappa_\perp$. It should also be noted that perpendicular diffusion does not necessarily describe crossing magnetic field lines.} The antisymmetric diffusion coefficients $\kappa_\mathrm{A}$ are often negligible or absorbed into potential drift terms \cite{kopp2012}. The exact form of the diffusion tensor $\hat{\kappa}$ strongly depends on the exact realization of the turbulent magnetic field, including its power spectrum $G(k)$ and the turbulence ratio $\eta=b/B$. Some of the parameters can be deduced by comparing transport model predictions to observational data, such as the boron-to-carbon ratio \cite{ams2014}, by numerical test particle simulations as presented here, and theoretical models \cite{Oughton2021}.

In particular, the so-called \textit{leaky box model} \cite{Cowsik1967}
of the Milky Way predicts that the cosmic-ray energy spectrum observed at Earth is steepened during propagation: Using the transport equation (\ref{eq:ParkerTransport}) and reducing it to a stationary case ($\partial n/\partial t \approx 0$) in which diffusion is the dominant process, we obtain a simplified equation
\begin{equation}
    0\approx -\frac{n}{\tau_{\rm diff}}+S\,.
\end{equation}
Here, we approximate the diffusion process via the time scale $\tau_{\rm diff}=d^2/\kappa$, defined via
\begin{equation}
-\frac{n}{\tau_{\rm diff}} \approx \nabla(\hat{\kappa}\nabla n)   \,,
\end{equation}
where $d$ is the escape distance and $\kappa$ is the diffusion scalar. Thus, the cosmic-ray density is given by the ratio of the source spectrum and the diffusion coefficient,
\begin{equation}
    n\propto \frac{S}{\kappa}\,.
\end{equation}
Assuming diffusive shock acceleration, one arrives at a power-law spectrum at injection, i.e.\ $S(E)\propto E^{-\gamma_{\rm s}}$. The diffusion coefficient in quasi-linear theory (QLT) with a wave spectrum that follows a power law also becomes a power law (see e.g.\ \cite{sigl2017}), $\kappa_i (E)\propto E^{\gamma_i}$ with $i=\,\parallel,\,\perp$. Thus, in case of diffusion-dominance, the energy spectrum of cosmic rays after propagation is steepened, i.e.\ $n(E)\propto E^{-\gamma_{\rm s}-\gamma_i}$ \cite{Berezinskii1990}. These arguments are based on QLT, in which only linear terms in the distortions of the electromagnetic fields and particle population with respect to the undisturbed fields are being considered to simplify the kinetic equations and to facilitate their analytical treatment. In particular, if we assume isotropic turbulence, we obtain an isotropic wave-vector spectrum of the form $G(k)\propto E^{-\beta}$, with $\beta = 5/3$ \cite{kolmogorov1941a} for Kolmogorov-like turbulence and $\beta = 3/2$ for the Kraichnan type \cite{Kraichnan1965}. In QLT, this leads to a parallel diffusion coefficient $\kappa_\parallel \propto E^{2-\beta}$ \cite{jokipii1966}. While both spectral indices describe the observed solar wind turbulence inertial range within the uncertainties \cite{Smith2006}, we only consider Kolmogorov turbulence in the following. Based on this QLT prediction, there is an expected difference of about 0.2 in the index of the energy dependence of the diffusion coefficients between the two turbulence models, which results in an underlying uncertainty in our results due to the uncertainties in the turbulence model.
 
Previous numerical simulations of particle transport in a combined turbulent field $b$ and homogeneous field $B$ were in agreement with this spectral index up to high $b/B \sim 1$ \cite{Giacalone1999, Casse2001, DeMarco2007}. Such simulations are typically performed on a 3-dimensional Cartesian grid with spacing $s$ and number of grid points $N_{\rm grid}$, injecting relativistic particles from a point source in a homogeneous $(\vec{B}$ plus turbulent $\vec{b}$ magnetic field in order to derive the diffusion coefficient. Radiative losses are not considered, since they are not relevant here.
Recent studies have, however, pointed out that these results need to be interpreted carefully, as the energy range in which the simulations are fully diffusive is strongly limited \cite{Minnie2007, lange2013, Snodin2015, Giacinti2017, Reichherzer2020, Dundovic2020}. 
Using simulations in the fully resonant scattering regime, we quantitatively investigate the energy behavior of the diffusion coefficient as a function of the turbulence ratio $b/B$. Finally, we interpret these results in the context of recent measurements of the diffuse cosmic-ray flux in the Milky Way.

\section{Simulations of the diffusion coefficients} \label{diffusion:sec}
Our simulations are based on the Taylor-Green-Kubo (TGK) formalism \cite{shalchi2009}, see e.g. \cite{Giacalone1999,Casse2001,Globus2007,DeMarco2007,Minnie2007, Snodin2015, Giacinti2017, Subedi2017}. Building on work that focused on specific parameter points and resolutions, we conduct a systematic simulation and analysis setup, similar to \cite{Reichherzer2020} that reveals key conditions on numerical simulation requirements. This method uses the fact that the fundamental solution of the diffusion problem, $\kappa_{ii} \Delta f(x_i,t)=\delta f(x_i,t)/\delta t$, is known to be a Gaussian whose width is described by the diffusion coefficient $\kappa_{ii}$. The second moment of the deviation in $x_i$ provides an analytical solution $\left<\Delta x_{i}^{2}\right>=2\,t\,\kappa_{ii}$.
The left-hand side of this equation can be calculated in simulations of particles that are emitted from a point source, which is placed in a field composed of a homogeneous component $\vec{B}=B\,\vec{e}_x$ and a turbulent component $\vec{b}$. Here, we use a Kolmogorov-type spectrum, i.e., isotropic and without intermittency
\begin{equation}
G(k) \propto 
\begin{cases} 
     0  & \text{if } k < k_\mathrm{min}\,,\\
   \left(\frac{k}{k_{\mathrm{min}}}\right)^{-\alpha}    & \text{if } k_\mathrm{min} \leq k \leq k_\mathrm{max}\,,\\
   0   & \text{if } k_\mathrm{max} < k\,,
  \end{cases}
  \label{kspectrum:equ}
\end{equation}
with $\alpha=5/3$, $k_{\min}=2\pi/l_{\max}$, and $k_{\max}=2\pi/l_{\min}$, where $l_\mathrm{min}$ is defined as the smallest numerically resolved wavelength, and $l_\mathrm{max}$ represents the largest wavelength used in the simulation. Evaluation of the correlation length in the limit $l_{\min}/l_{\max}\ll 1$ yields $l_\mathrm{c} \approx l_{\max}/5$ \cite{Harari2013}. The synthetic isotropic three-dimensional turbulent magnetic field is generated and stored discretely on a regular, three-dimensional Cartesian grid with $N_\mathrm{grid}^3$ grid points and isotropic spacing $s_{\mathrm{spacing}}$ using the inverse Fourier transform of field vectors $\mathbf{{b}}(\mathbf{{k}})$ that are computed on a regular grid in three-dimensional wavenumber space.
Linear interpolation yields the magnetic field at an arbitrary trajectory position between grid points.

For discrete step sizes $s_\mathrm{step} = v \, \Delta t$, the diffusion coefficient can be calculated as
\begin{equation}
\begin{split}
    \kappa_{ii} (t) & = \sum_{n = 0}^{t/(\Delta t)} \Delta t \left\langle v_i(n\Delta t)v_i(0) \right\rangle \\& = 
    \sum_{n = 0}^{t/(\Delta t)} \left\langle v_i(n\Delta t)\Delta x_i(0) \right\rangle.
\end{split}
\end{equation}
$i$ indicates the three spatial directions. Here, we consider parallel diffusion $\kappa_{\parallel}$, as well as perpendicular diffusion $\kappa_{\perp}$.
Furthermore, $t=n \, \Delta t$ is the time after $n$ time steps. We propagate particles on a grid with $N_\mathrm{grid}=1024$, $s=0.85$~pc, $l_{\min}=1.7$~pc, $l_{\max}=82.45$~pc, and a step size $s_{\mathrm{step}}=\min(r_\mathrm{g}/5,l_\mathrm{max}/20)$ to ensure that the gyration motions as well as the fluctuations are sufficiently resolved. With the large grid and correspondingly broad spectrum, particles always find waves for interactions \cite{schlegel2019}. The comparatively large $l_\mathrm{grid}=N_\mathrm{grid} \cdot s \approx 50 l_\mathrm{c}$ is chosen in order to reduce problems caused by the continuation of the turbulence at the grid-box boundaries.

While we adopt inertial range Kolmogorov scaling for the turbulence, this is an idealization. Interstellar turbulence is driven on many scales, from as much as 100 pc by superbubbles to kinetic scales by cosmic rays themselves \cite{Lazarian2015}. Moreover, Alfvénic turbulence, unlike hydrodynamic Kolmogorov turbulence is known to be highly anisotropic. While compressibility effects can generate an isotropic component \cite{Lazarian2015}, this mechanism is unlikely to result in a simple inertial cascade. Whereas the turbulence-dependent energy scaling, which will be determined in the next section, can be scaled from the large scales considered here to small scales straightforwardly, the drivers of turbulence relevant on every scale considered must be taken into account for the normalization of the diffusion coefficients.\\\\
For Gaussian particle distributions, the running diffusion coefficient $\kappa(t)$ is expected to converge in time to a constant value $\kappa(t)\rightarrow \kappa$. Simulations are stopped after several orders of gyrations once the running diffusion coefficients converge, and the final value of the diffusion coefficient is taken as the steady-state one. We repeat these simulations several times with 2000 particles and the same parameters but varying random numbers for turbulence generation. As the low turbulence levels in the RSR require more statistics, we use 50 random seeds in the RSR for $b/B \leq 2$ and otherwise 20 random seeds. We perform this step for different reduced rigidities and fit a power law $\kappa\propto \rho^{\gamma}$ to the RSR, where particles experience resonant scatterings: $l_{\min}/(\pi\,l_c(b/B))<\rho<l_{\max}/(2\pi\,l_c)$. We determine the reduced rigidity scaling for a broad range of turbulence levels $0.05\lesssim b/B \lesssim 10$, while $B=1\mu \mathrm{G}$ remains constant. 

\section{Simulation results}\label{sec:4}
Over the last decades there have been great advances in direct numerical calculations of particle transport in mean magnetic fields subject to turbulent perturbations with a Kolmogorov-like spectrum accompanied by the adaptation of theoretical models to match these simulation results. In addition to studies of parallel diffusion coefficients, the perpendicular diffusion coefficients and the relationship between the two components have been the subject of research \cite{Giacalone1999, Mace2000, Casse2001, Matthaeus2003, Candia2004, DeMarco2007, Minnie2007, Fatuzzo2010, Plotnikov2011, Harari2013, Harari2015, Snodin2015, Subedi2017, Giacinti2017, Reichherzer2020, Dundovic2020}. Good agreement between analytical models such as QLT (in some regimes), nonlinear guiding center theory and unified nonlinear theory with test particle simulations was found for two-dimensional, slab, or composite (two-dimensional \& slab) turbulence (see e.g.~\cite{Shalchi2020SSRv} for a review). Despite these advances, for isotropic three-dimensional turbulence, tension with these theories was found at small reduced rigidities \cite{Reichherzer2020, Dundovic2020}.

\begin{figure}
\includegraphics[width=\linewidth]{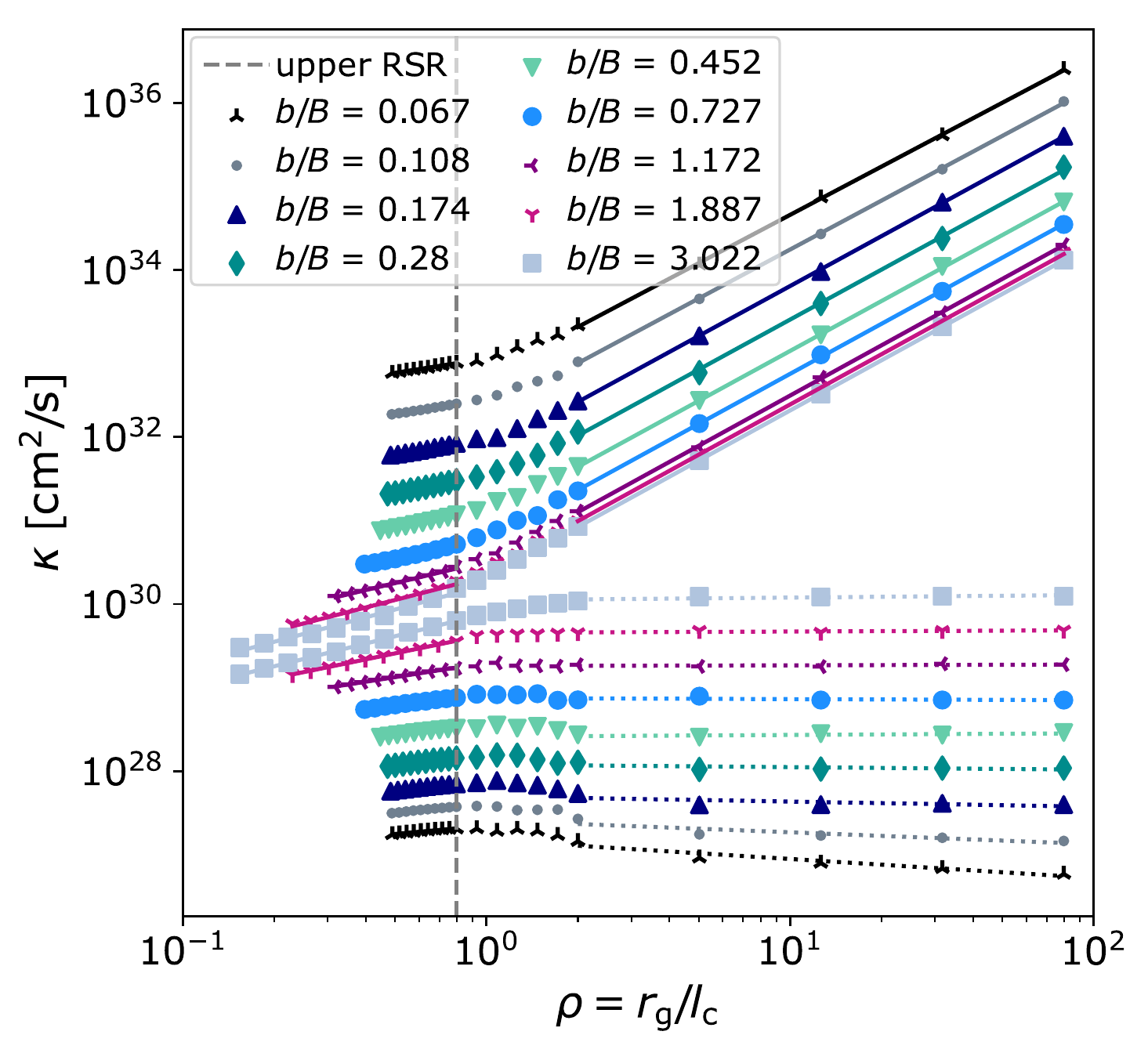}
\caption{Parallel and perpendicular diffusion coefficients as functions of $\rho$ and $E$ for different turbulence levels. Only simulation results above the lower boundary of the RSR $\rho \gtrsim l_\mathrm{min}/(\pi(b/B)l_\mathrm{c})$, above $E \gtrsim 7.5$~PeV, and below the upper boundary of the RSR $\rho \lesssim 5/(2\pi)$ are considered for determining the energy scaling of $\kappa_\parallel$ and $\kappa_\perp$ within RSR. Fits are performed to these simulated diffusion coefficients.
The parameters are: $l_\mathrm{c} = 17.8$ pc, $l_\mathrm{min} = 1.7$ pc, $l_\mathrm{max} = 82.45$ pc, $s = 0.85$ pc, $N_\mathrm{grid}$ = 1024. Each presented data point is the mean of 20-50 diffusion coefficients, each simulated with the same parameters but with a different turbulent field realization. The decreasing range of the RSR for smaller $b/B$ leads to an increasing error in the slopes of the fits.\label{fig:fig3}}
\end{figure}

In Fig.~\ref{fig:fig3} we show simulation results of the parallel (solid lines) and perpendicular (dotted lines) diffusion coefficients calculated using 2000 particles in each simulation. For each of these ratios, 20 different energies are simulated. 
The statistical uncertainties of the diffusion coefficients are calculated by repeating each simulation 20-50 times with different realizations of the turbulent magnetic field. The averaged diffusion coefficients are shown as functions of the reduced rigidity $\rho$. The statistical uncertainties are too small to be visible.\\\\
The classification of the diffusive transport into different transport regimes based on the particle energy and the turbulence level, and especially the definition of the resonant scattering regime (RSR), allows the comparison of simulated data with the QLT \cite{Reichherzer2020}. This is due to the fact that only in the RSR the conditions of QLT for resonant scattering over the whole pitch angle range are met in the simulations. As the lower RSR boundary $\rho = l_\mathrm{min}/(\pi l_\mathrm{c} b/B)$ utilizes the approximation of small turbulence levels, we apply this formula to determine the lower RSR boundary of the lowest used turbulence level $b/B = 0.067$. Since this lower boundary provides the smallest possible error from the underlying approximation for small turbulence levels, we also use the corresponding energy $E=7.5$~PeV for all other turbulence levels as the lower boundary. While this approach does not exploit the full width of the RSR, as the RSR extends to smaller reduced rigidities as the turbulence level increases, the approximation of small angles still applies to all fits. The small errors in the fits are an indication that our chosen range in reduced rigidity is sufficiently large. The upper RSR boundary is $\rho = 5/(2\pi)$ for all turbulence levels.\\\\
Note that the simulated diffusion coefficients shown in Fig.~\ref{fig:fig3} show qualitative similarities with comparable studies for isotropic three-dimensional turbulence (see e.g.~\cite{Snodin2015}), but also with two-dimensional turbulence for the perpendicular component at large reduced rigidities \cite{Dempers2020}.\\\\
In the following, first the parallel and then the perpendicular diffusion coefficients are investigated in the RSR and the QBR.

\subsection{Parallel component}
\begin{figure}
\includegraphics[width=\linewidth]{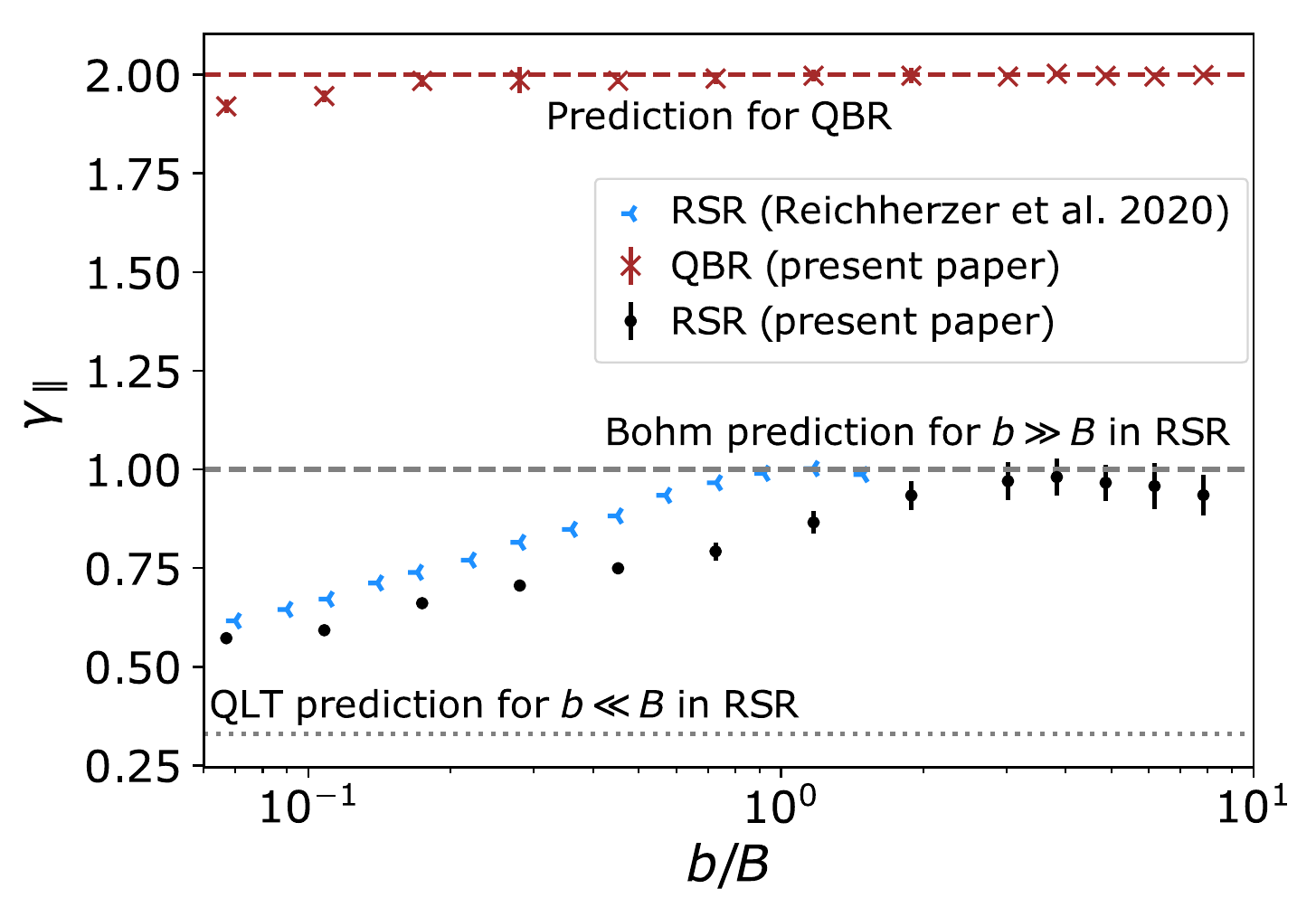}%
\caption{Turbulence-level-dependent spectral index of the parallel diffusion coefficients in the RSR and QBR with statistical uncertainties. The simulated diffusion coefficients are fitted linearly in the log-log representation for each ratio of $b/B$ as shown in Fig.~\ref{fig:fig3}, with the slopes shown in this plot. The simulation
parameters are: $l_\mathrm{min} = 1.7$ pc, $l_\mathrm{max} = 82.45$ pc, $l_\mathrm{c} = 17.8$ pc, $s = 0.85$ pc, $N_\mathrm{grid}$ = 1024.\label{fig:fig4}}
\end{figure}

Especially for lower turbulence levels, the parallel diffusion coefficients are several orders of magnitude larger than the perpendicular components, as presented in Fig.~\ref{fig:fig3}. This is due to the different scaling with the turbulence level. While the parallel component scales with $(b/B)^{-2}$, the perpendicular component decreases with decreasing turbulence level. Analytical theories describe the scaling of the turbulence dependence of the diffusion coefficients (see e.g.\,\cite{jokipii1966, shalchi2009}).\\
We perform fits in the RSR and QBR to determine the energy scaling of the diffusion coefficients for different ratios of $b/B$, as theory predicts a proportionality between $\log{\kappa}$ and $\log\rho$ in both regimes. Note that due to the limitations of the fluctuation range of the synthetic turbulence, we can perform the fits of the predicted power law in the RSR only over a relatively small energy range, however, while strictly following our derived boundaries of this energy regime.\\\\
The calculated power-law indices $\gamma_\parallel$ of the fits $\kappa_\parallel \propto \rho^{\gamma_\parallel}$ are presented for different turbulence levels in Fig.~\ref{fig:fig4}. The simulation results for the QBR are shown as red crosses. The $\gamma_\parallel$ values are in agreement with theoretical predictions of $\gamma_\parallel = 2$ \cite{Aloisio2004, Subedi2017}.\\\\
For the RSR, we compare our results with simulation results from \cite{Reichherzer2020}. The underlying simulation setups of the shown simulation results from \cite{Reichherzer2020} and this publication differ only in the spacing of the grid points for the generation of turbulence. While in \cite{Reichherzer2020}, a grid spacing of $l_\mathrm{min}/10$ was chosen, here we use the maximum possible grid spacing of $l_\mathrm{min}/2$, which is based on the sampling theorem. The grid size has a direct influence on two numerical effects: Smaller spacing between grid points improves the resolution of small scales and thus reduces the numerical errors when interpolating the magnetic field between surrounding grid points. And, a larger grid reduces the need to periodically continue the original box when particles diffuse spatially across the boundaries. A comparison with a gridless method for turbulence generation has provisionally shown good agreement with the larger grid spacing \cite{Schlegel2021}. The gridless method is based on \cite{Giacalone1999, Tautz2013, schlegel2019} and evaluates at each point in space the sum of pre-generated wavemodes while eliminating the need of storing discrete field vectors on the grid.\\\\
While the results of both setups are slightly shifted against each other, they follow the same trend: The QLT limit of $\gamma=1/3$ is not yet reached at the 5\% turbulence level, but is expected at even lower $b/B$ based on the trend. At $b/B\approx 2$, the upper limit for diffusion is reached, consistent with Bohm diffusion $\kappa_\parallel \propto \rho$.\\
Here we would like to mention that for the fits at high turbulence levels an increasing number of simulated diffusion coefficients at low rigidities are considered, as shown in Fig.~\ref{fig:fig3}. Towards lower $\rho$ values and thus smaller scales on which the particles gyrate, numerical effects such as interpolation errors increase, which is due to the limited resolution of the discrete grid on which the turbulence is stored, which is why the $\gamma$ values towards high turbulence levels may contain an additional systematic error, which is not indicated, but may lead to the deviation from the expected $\gamma \approx 1$.

\subsection{Perpendicular component}
\begin{figure}
\includegraphics[width=\linewidth]{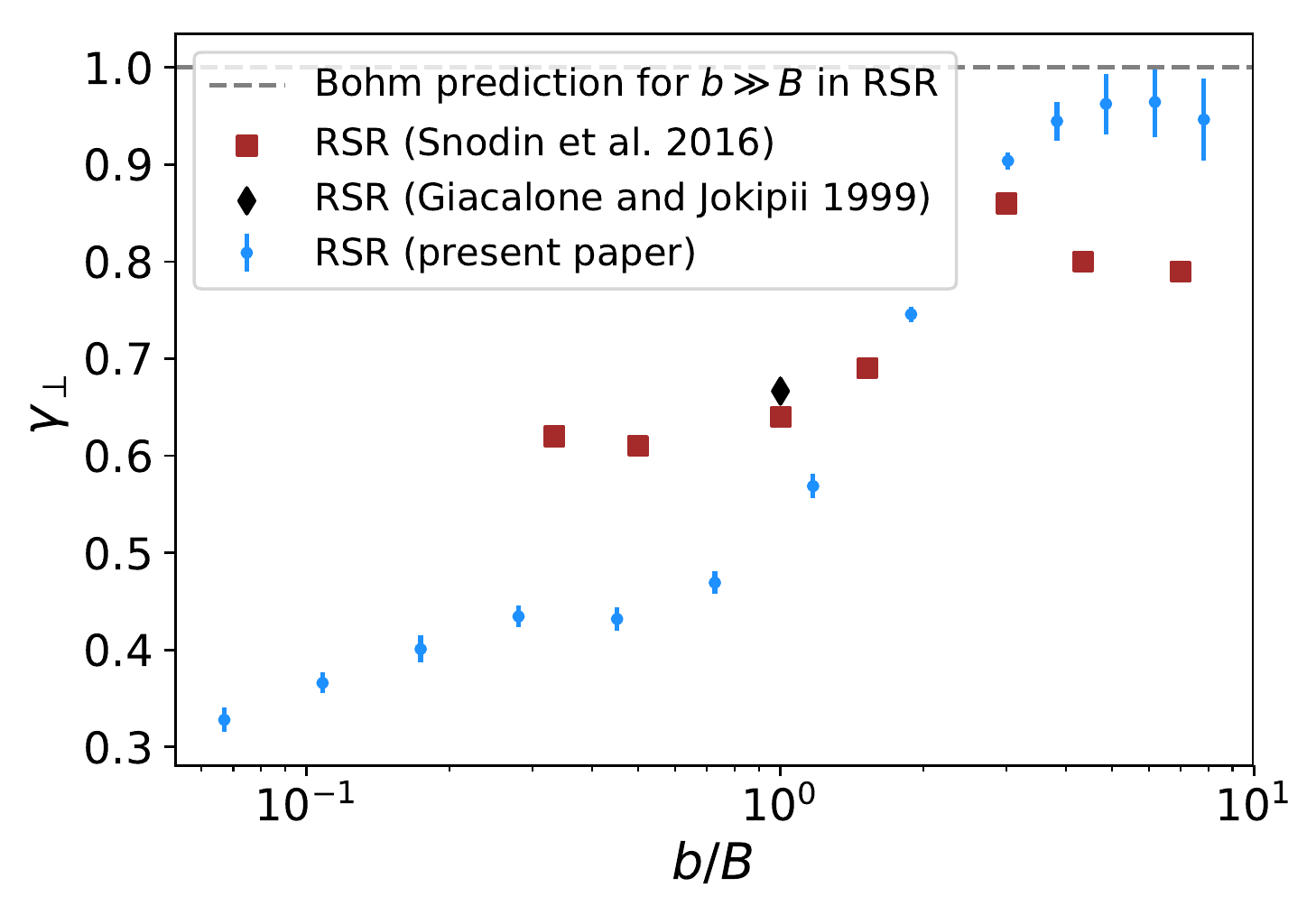}%
\caption{Turbulence-level-dependent spectral index of the perpendicular diffusion coefficients in the RSR and QBR with statistical uncertainties. The simulated diffusion coefficients are fitted linearly in the log-log representation for each ratio of $b/B$ as shown in Fig.~\ref{fig:fig3}, with the slopes shown in this plot. The simulation parameters are: $l_\mathrm{min} = 1.7$ pc, $l_\mathrm{max} = 82.45$ pc, $l_\mathrm{c} = 17.8$ pc, $s = 0.85$ pc, $N_\mathrm{grid}$ = 1024. Simulations from the literature \cite{Giacalone1999, Snodin2015, Snodin_privcom} and (A. Snodin, personal communication, 2020) are also shown for comparison. It should be noted that \cite{Giacalone1999} assumes non-relativistic particles.\label{fig:fig5}}
\end{figure}

\begin{figure}
\includegraphics[width=\linewidth]{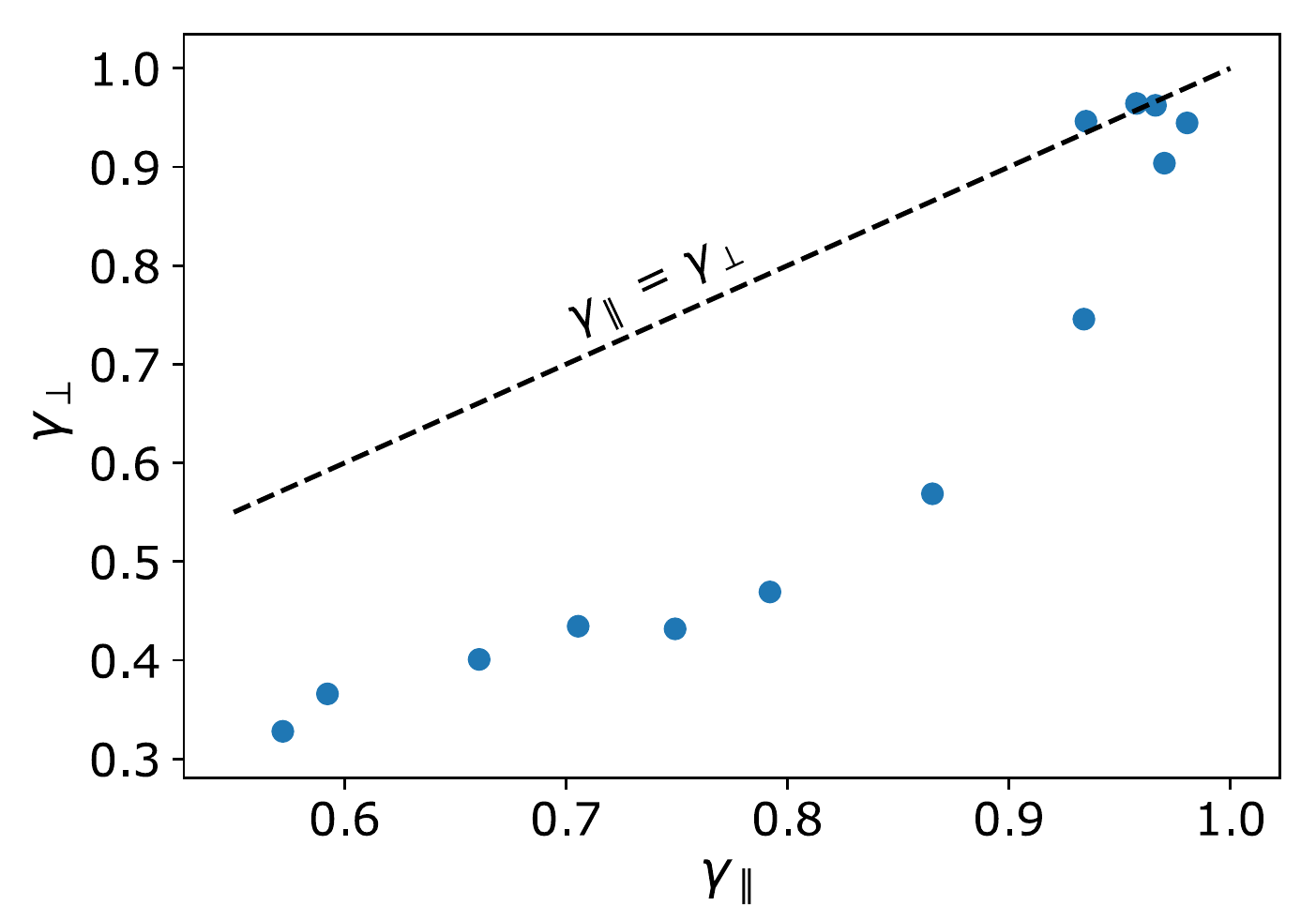}%
\caption{Spectral index of the perpendicular diffusion coefficient as a function of the spectral index of the parallel diffusion coefficient, as shown in Figs. \ref{fig:fig4} and \ref{fig:fig5}. The turbulence levels increase from the lower left to the upper right. The black dashed line is a reference for the case of equal indices along the perpendicular and parallel components. Only for high turbulence levels, the simulation values are close to this reference line. \label{fig:fig6}}
\end{figure}

Fits based on $\kappa_\perp \propto \rho^{\gamma_\perp}$ for the perpendicular components are shown in Fig.~\ref{fig:fig5}. For comparison, simulated values from literature are included. 
The slopes are quite sensitive to the range of $\rho$ considered as pointed out in \cite{Snodin2015, Reichherzer2020}, which is why the fitted slopes vary based on different fit ranges in different publications. Nevertheless, the same trend can be seen across all publications: Increasing turbulence levels lead to a coincidence of the parallel and perpendicular components, as $\gamma_\parallel$ and $\gamma_\perp$ converge towards 1\footnote{Equivalently to the parallel diffusion coefficient, the small systematic deviation from the value of $\gamma=1$ can be explained by numerical effects due to the limited resolution of the turbulent magnetic field and the associated interpolation effects by considering diffusion coefficients at low reduced rigidities}. This is expected since the directional component of the background field loses influence with increasing turbulence levels and diffusion becomes isotropic. As with parallel diffusion, Bohm diffusion must also occur for perpendicular diffusion with regards to the background field for large turbulence levels (Fig.~\ref{fig:fig6}).\\
At lower turbulence levels, however, the increase in field-line random walk (FLRW) is expected, since the diffusion coefficient of the field lines scales proportionally to $(b/B)^2$. Since FLRW is only independent of energy when particles follow the field lines, there is a transition in behavior at roughly the energy at which the field lines separate by more than one gyroradius over the distance the particles travel in one gyroorbit. The observable decrease of the energy dependence of particle diffusion with decreasing turbulence levels, which is associated with a smaller value of $\gamma_\perp$, can thus be explained by the increasing importance of particles diffusing through FLRW.

\section{Discussion in the astrophysical context}

What do our results imply in the context of astrophysics? In general, we can show that the diffusion tensor strongly depends on the turbulence level --- a fact that is typically not taken into account in astrophysical simulations. A prominent example where this $b/B$-dependence can play a role is the high-energy signatures for our own Galaxy, where the Fermi satellite has measured gamma-ray emission at GeV energies \cite{Acero2016, Yang2016}. It was possible to derive the contribution from hadrons and thereby quantify both the cosmic-ray proton number density and the index of the cosmic-ray proton spectrum as a function of galactocentric radius. These results are presented in panel a) of Fig.\ \ref{fig:multipanel}. From these data follows what is known as the \textit{cosmic-ray gradient problem}. In particular, there is a gradient in the proton index, ranging from relatively flat spectra in the innermost $\sim 5$~pc of the Galaxy ($E^{-2.3}-E^{-2.5}$) to a steep index of close to $E^{-3.0}$ at the outermost radii. Recent interpretations of this feature include phenomenological models of a change in the diffusion index based on geometric effects of anisotropic diffusion \cite{Evoli2012}, a galactocentric radius dependent diffusion tensor \cite{dragon, gaggero2015, gaggero_prl2017, Cerri2017}, and nonlinear cosmic ray transport with scattering and advection off self-generated turbulence in combination with galactocentric dependent cosmic ray source distributions \cite{Recchia2016}. \\\\
With the turbulence-dependent energy dependence of diffusion presented here, we can expand the anisotropic diffusion models presented in the literature.
To simplify the transport model even further we neglect momentum diffusion $\kappa_{pp}=0$ and adiabatic effects $\nabla\cdot\vec{u}=0$ and assume a stationary state (see Eq. \ref{eq:ParkerTransport})
\begin{align}
    S + \nabla\left(\hat{\kappa}\,\nabla n \right) -\vec{u}\cdot\nabla n  =  \frac{\partial n}{\partial t}\approx 0\, .
\end{align}
As the number of individual sources that contribute to $S$ change with the position in the Galaxy, so does $S$. This does, however, not change the spectral index systematically, as in diffusive shock acceleration, the spectral index mainly depends on the strength of the shock. Only the intensity of the signal changes as the source density varies with galactocentric radius. The diffusion term can be approximated by using the effective escape distance $d_{\parallel}$ and $d_{\perp}$ in the parallel and perpendicular directions, respectively, as the spatial dependence of the term
\begin{equation}
    \nabla\left(\hat{\kappa}\,\nabla n \right)\approx \left( \frac{\kappa_{\parallel}}{d_{\parallel}^{2}}+\frac{\kappa_{\perp}}{d_{\perp}^2}\right)n=-\left(\frac{n}{\tau_{\parallel}}+\frac{n}{\tau_{\perp}}\right)\sim -\frac{n}{\tau_{\rm diff}}.
\end{equation}
Here, the factors have been identified by escape times $\tau_i \equiv d_{i}^{2}/\kappa_i$, with $i=\,\parallel,\,\perp$. This is a very rough simplification, neglecting the full radial dependence of the diffusion tensor. We are doing this as, at this point, we simply want to discuss the effect caused by the perpendicular and parallel components and interpret the escape times as radially dependent for now. A quantitative model will have to take into account the full spatial structure of these terms.\\\\ 

In our model, the central question is: which of these two time scales will dominate at what radius? The diffusion time scale is given by the shorter time scale, $\tau_{\rm diff}\sim \min\left(\tau_{\parallel},\,\tau_{\perp} \right)$.

\begin{figure}
\includegraphics[width=\linewidth]{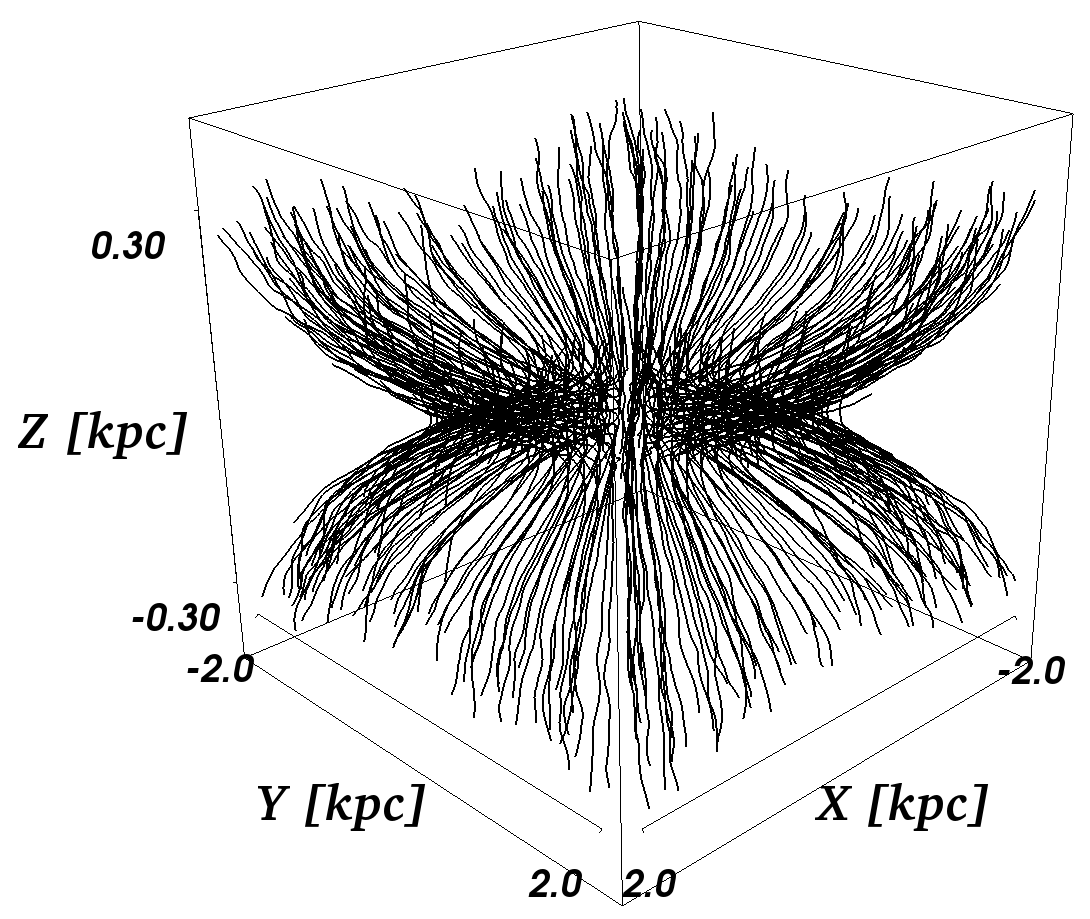}%
\caption{Three-dimensional view on the magnetic field lines in the Galactic center, using a combination of the global field \cite{Kleimann2019} and an in-plane Galactic center component \cite{Guenduez2020}. \label{fig:3dimfield_center}}
\end{figure}

\begin{figure}
\includegraphics[width=\linewidth]{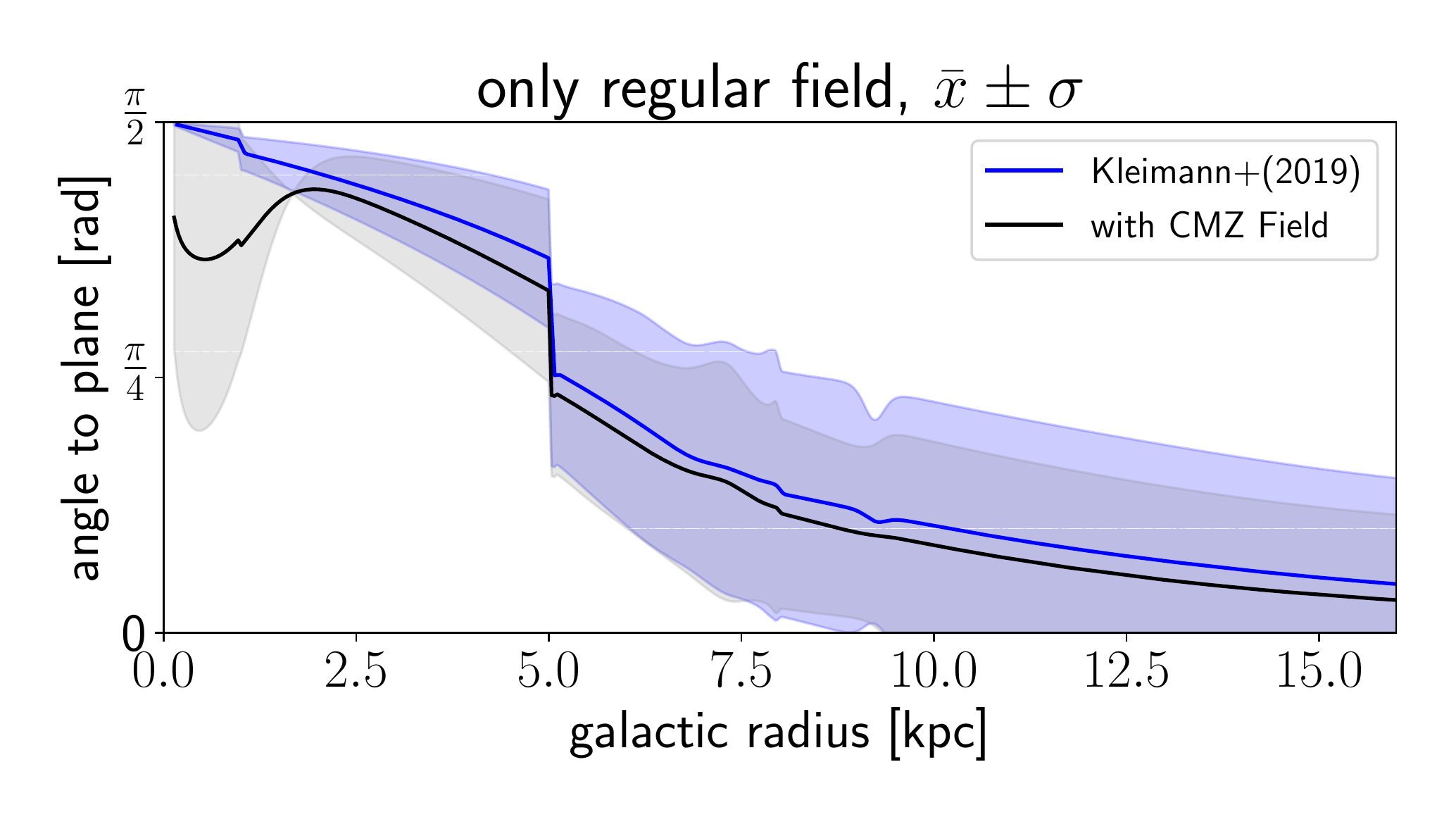}%
\caption{Mean angle of the field lines with respect to the Galactic plane as a function of the galactocentric radius. The blue line shows the pure global field \cite{Kleimann2019}, the black line shows the combination of \cite{Kleimann2019} with the Galactic Center field model of \cite{Guenduez2020}.\label{fig:angle_mean}}
\end{figure}

\begin{figure}
\includegraphics[width=\linewidth]{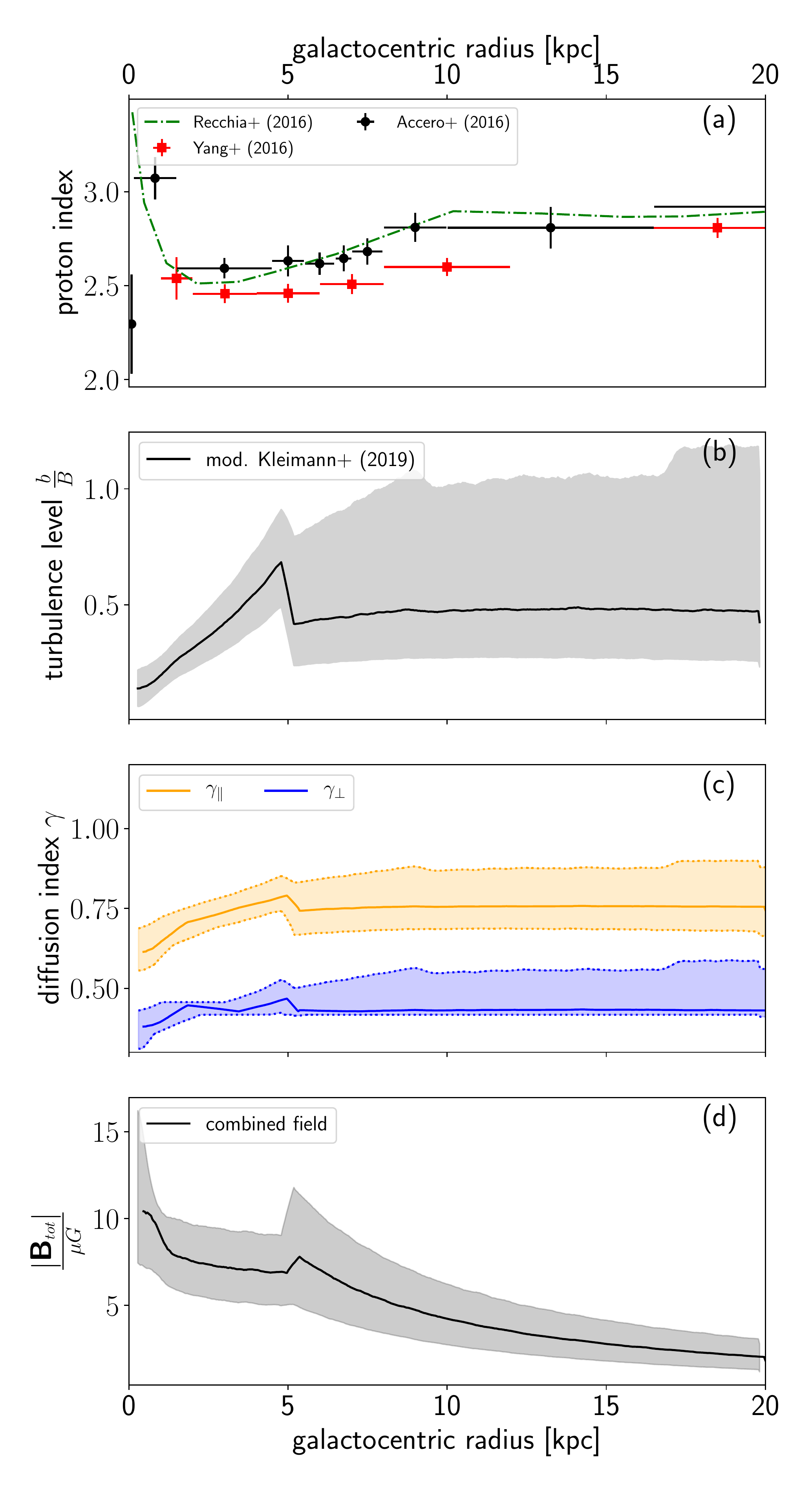}%
\caption{Different properties of the Galaxy as a function of galactocentric radius, starting from the top: (a) cosmic-ray spectral index at GeV energies, derived from Fermi measurements \cite{Acero2016, Yang2016, Recchia2016}; (b) the displayed turbulence level is derived from considering the regular component of the model described in \cite{Kleimann2019} and the turbulent component from \cite{JF12}. (Comment R2 1.) We have downscaled the latter component by a factor of 10 for more realistic values}; (c) perpendicular and parallel diffusion indices as derived at a fixed level of $b/B$ (this work); (d) total magnetic field strength of the combined regular field of \cite{Kleimann2019} and \cite{Guenduez2020}. 
\label{fig:multipanel}
\end{figure}

For the two individual diffusion time scales, these dependencies are as follows: 
\begin{itemize}
    \item \textbf{Energy and $b/B$:} As discussed in the earlier sections, the energy dependence of the two coefficients is
    \begin{equation}
        \kappa_i\propto \left(\frac{b}{B}\right)^{\pm 2} \rho^{\gamma_i}\propto \left(\frac{b}{B}\right)^{\pm 2}\left(\frac{E}{B_{\rm tot}}\right)^{\gamma_i}
    \end{equation} 
    Here, the dependence is $(b/B)^{\pm 2}$ with a negative and positive turbulence level index for parallel and perpendicular diffusion, respectively. Thus, as the time scales behave as $\kappa_i^{-1}$, we have the following dependences for the parallel and perpendicular time scales:
    \begin{eqnarray}
        \tau_{\parallel}&\propto& \left(\frac{b}{B}\right)^{2}\,B_{\rm tot}^{\gamma_\parallel}\,E^{-\gamma_\parallel},\\
        \tau_{\perp}&\propto& \left(\frac{b}{B}\right)^{-2}\,B_{\rm tot}^{\gamma_\perp}\,E^{-\gamma_\perp}\,.
    \end{eqnarray}
    \item \textbf{Field direction/galactocentric distance:} The escape distance for diffusive transport depends on the field direction, which changes with galactocentric distance: for parallel transport, the escape direction is along the field lines. Figure \ref{fig:3dimfield_center} shows the magnetic field in the Galactic center region (central $2$~kpc with a height of $300$~pc). The field is a combination of the global field first presented by \cite{JF12}, here used in a modified version \cite{Kleimann2019}, plus a component in the galactic plane presented in \cite{Guenduez2020}. This combination of fields is necessary, whereas global field models omit the in-plane component in the center. It can be seen here that even with the in-plane component, the field lines are essentially pointing perpendicular to the Galactic plane in the Galactic center region. This can also be seen in Fig.\ \ref{fig:angle_mean}, where the mean angle of the field lines at a given galactocentric radius with respect to the plane direction is shown. Up to radii around $3-5$~kpc, the angle is very close to $90^{\circ}$, while at values larger than $r_{\rm gc}\sim 5$~kpc, it becomes significantly smaller than $45^{\circ}$. This is consistent with the presence of a Galactic wind in the inner $3-5$~kpc \cite{Everett2008, Everett2010}, which can contribute to orient the field lines in the direction of the wind speed, which is perpendicular to the plane.

    In the outer region at $r_{\rm gc}>5$~kpc, escape via parallel diffusion therefore preferentially occurs along the plane. The relevant escape distance for parallel transport is therefore a function of galactocentric radius. As a simplification, the parallel escape distance, defined as the length of the escape path in parallel along the mean magnetic field, can be approximated as the scale height $d_{\parallel}\approx H\approx 300$~pc in the inner Galaxy, i.e.\ for $r_{\rm gc}\lesssim 5$~kpc. In the outer Galaxy, $r_{\rm gc}\gtrsim 5$~kpc, it can be approximated as the in-plane propagation distance. The latter is significantly longer, $d_{\parallel}> r_{\max}-r_{\rm gc} \gg 300$~pc for $r_{\rm gc}>5$~kpc, with $r_{\max}\sim 20$~kpc, especially when considering that particles will not diffuse out in straight lines along $\vec{e}_r$, but rather follow the field lines.  
    The perpendicular transport shows the opposite behavior: in the inner Galaxy, \textit{perpendicular} to the field lines means in-plane propagation and thus $d_{\perp}> r_{\max}-r_\mathrm{gc}\gg 300$~pc. In the outer Galaxy, perpendicular transport approaches an orientation perpendicular to the plane, with $d_{\perp}\approx H\approx 300$~pc. A larger escape distance is accompanied by a larger escape time, which leads to the fact that in the center, parallel transport results in the fastest escape of particles. In the outer Galaxy, perpendicular transport is accompanied by shorter escape distances \textit{until} the escape distance along the plane in the outermost part of the Galaxy becomes comparable to the scale height. Thus, at the outskirts of the Galaxy, parallel transport will dominate again. 
\end{itemize}

Combining these two effects, we get the diffusion time scale
\begin{align}
    \tau_{\rm diff}(r_{\rm gc})\approx \left\{\begin{array}{ll}
    \tau_{\parallel}\propto \left(\frac{b}{B}\right)^{2}\left(\frac{E}{B_{\rm tot}}\right)^{-\gamma_\parallel}\left\langle\frac{d_{\parallel}^2}{H^2}\right\rangle&\rm{for}~r_{\rm gc}\lesssim 5~\rm{kpc}\\
    \tau_{\perp}\propto \left(\frac{b}{B}\right)^{-2}\left(\frac{E}{B_{\rm tot}}\right)^{-\gamma_\perp}\left\langle\frac{d_{\perp}^2}{H^2}\right\rangle&\rm{elsewhere}\\
    \tau_{\parallel}\propto \left(\frac{b}{B}\right)^{2}\left(\frac{E}{B_{\rm tot}}\right)^{-\gamma_\parallel}\left\langle\frac{d_{\parallel}^2}{H^2}\right\rangle&\rm{for}~r_{\rm gc}\gtrsim 19~\rm{kpc}
    \end{array}\right.
\end{align}
Here, the notation $\langle...\rangle$ refers to averaging over all field lines at a given galactocentric radius. It should be noted that $d_{\parallel}(r)$ and $d_{\perp}(r)$ represent functions at a fixed position in the Galaxy and averaging over the scale height $ z \in [-H,+H]$ needs to be performed for detailed results. We refrain from doing so at this point because we would like to keep the argument simple and these details are not expected to change the interpretation. 

Note that inhomogeneities in $B$ at the $\mathrm{kpc}$ scale are apparent in the Galaxy. This implies that drifts will occur with characteristic escape time scales on the order of $\tau_\mathrm{drift} \sim 10^{13}\cdot(E/100\,\mathrm{GeV})^{-1}$\,yr based on drift velocities \cite{Strauss2012} along the $z$-direction within the setup of our toy model. While a complete quantitative picture of cosmic-ray diffusion must necessarily incorporate such drift physics, the conclusions at which the present work has arrived can still be expected to apply, given that these drifts cause bulk rather than diffusive motion and their time scales are many orders larger than those of diffusion and advection derived in the following.\\\\

We can now design the following toy model to investigate the dependence on the galactocentric radius: based on the parameters that are displayed in Fig.\ \ref{fig:multipanel}, there is a change in the physics of the turbulence at $r_0\sim 5$~kpc that should be reflected in the equations. Here, we use a simple ansatz based on the turbulence level shown in Fig.\ \ref{fig:multipanel}:
\begin{equation}
    \frac{b}{B}=\left\{\begin{array}{cc}
    \frac{r_{\rm gc}}{r_0}     & (r_{\rm gc}<r_0) \\
    1     & (r_{\rm gc}>r_0) 
    \end{array} \right.\,.
\end{equation}
We can also parametrize the total magnetic field in a similar way,
\begin{equation}
    B_{\rm tot}\propto \left(\frac{r_{\rm gc}}{r_0}\right)^{-\beta}\,.
\end{equation}
Here, we roughly approximate the total magnetic field behavior with $r_{\rm gc}$ based on the function shown in Fig.~\ref{fig:multipanel} and use $\beta\sim 0.1$ in the inner $5$~kpc and $\beta\sim 1$ for $r_{\rm gc}>5$~kpc. For the diffusion indices, we use constant values, as the dependency is rather small for the individual indices,
\begin{align}
\gamma_{\parallel}&\sim 0.7\label{eq:gamma_parallel}\\
\gamma_{\perp} &\sim 0.4\label{eq:gamma_perp}\,.
\end{align}
We approximate the escape distances with the scale height for $r<19$~kpc, since that is the dominant escape direction. For larger radii, $d_\parallel$ depends on the position in the Galaxy.
\begin{align}
    \tau_{\rm diff}(r_{\rm gc})\propto\left\{\begin{array}{ll}
     \left(\frac{r_{\rm gc}}{r_0}\right)^{2-\beta\,\gamma_\parallel}\,E^{-\gamma_\parallel}&\rm{for}~r_{\rm gc}\lesssim 5~\rm{kpc}\\
     \left(\frac{r_{\rm gc}}{r_0}\right)^{-\beta\,\gamma_\perp}\,E^{-\gamma_\perp}&\rm{elsewhere}\\
     \left(\frac{r_{\rm gc}}{r_0}\right)^{-\beta\,\gamma_\parallel}\,E^{-\gamma_\parallel}\,\left\langle\frac{d_{\parallel}^2}{H^2}\right\rangle&\rm{for}~r_{\rm gc}\gtrsim 19~\rm{kpc}
    \end{array}\right.
\end{align}

This result implies a somewhat steeper energy spectrum in the inner Galaxy, $n\propto E^{-\gamma_{\rm s}-\gamma_i} \sim E^{-2.9}$, using $\gamma_{\rm s} \sim 2.2 - 2.4$ and $\gamma_\parallel\sim 0.7$ as compared to the outer Galaxy, where perpendicular transport dominates ($\gamma_\perp \sim 0.4$) and thus $n\propto E^{-2.7\pm0.1}$. needed. Here, we want to emphasize that the measurements of diffusive emission in the Galactic center (see the two innermost data points in Fig.~\ref{fig:multipanel}) are dominated by contributions from point sources and are further confounded by interpolation effects during the analysis process \cite{Acero2016}, which limits the constraints imposed by these measurements in the Galactic center. However, as the observed change in the spectrum is from a rather flat one in the inner Galaxy, $n\propto E^{-2.3}$, to a steeper index in the outer region, $n\propto E^{-2.8\pm0.1}$, other effects need to play a role in the inner Galaxy, while the steepening in the outer Galaxy could be due to a change to diffusive behavior in general. Even a change from perpendicular to parallel escape in the outermost part of the Galaxy could be relevant. 

In the inner part of the Galaxy, a wind could play an important role as already discussed in \cite{dragon, Everett2008, Everett2010}. We also approximate the convective term in terms of a loss time scale, via
\begin{equation}
\vec{u}\cdot\nabla\,n\approx \frac{u}{d_{\rm conv}}\,n =\frac{1}{\tau_{\rm conv}}\,n\,,
\end{equation}
introducing the convection time scale \begin{equation}
\tau_{\rm    conv}=\frac{d_{\rm conv}}{u}\,.
\end{equation} 
It was shown in \cite{Everett2008, Everett2010} that a wind with speeds around $500-700$~km/s is present in the Galaxy.
Using an approximate value of  $u\sim 5\cdot  10^{7}$~cm/s and a scale height for the wind of $d_{\rm conv}\sim 300$~pc, we obtain a time scale of
\begin{equation}
    \tau_{\rm conv}\sim 5\cdot10^{5}~\rm{yr}\,,
\end{equation}
independent of the particle energy.

Here, the terminal wind velocity is used, which might be overestimating the actual wind velocity below $z<300$~pc. A dedicated wind model based on a stratified disk instead of a flat plane would improve the results.\\
We can compare this time scale with the diffusive one in the inner Galaxy,
\begin{equation}
    \tau_{\rm diff}\propto  \left(\frac{r_{\rm gc}}{r_0}\right)^{2-\beta\,\gamma_\parallel}\,\left(\frac{E}{E_0}\right)^{-\gamma_\parallel}\,.
\end{equation}
Here, we use the scaling with energy from our simulation findings (see Eqs.~(\ref{eq:gamma_parallel}), (\ref{eq:gamma_perp}), and surrounding text). However, the simulated absolute diffusion coefficients of the particles from Fig.~\ref{fig:fig3} would have to be extrapolated over several orders of magnitude to apply to our model according to the above formula. Note that the exact scaling depends on the cosmic ray energy and the galactocentric dependent magnetic field properties such as the field strenght and the correlation length of the turbulence. Due to our uncertainties for $\gamma$ from the fits, the errors in the scaling over several orders of magnitude would become too large. Therefore, we normalize the time scale $\tau_\mathrm{diff}(r_0=5\,\mathrm{kpc}, E=10\,\mathrm{GeV})\sim 10^{7}$~yr by using $\tau_\mathrm{diff}(r_\mathrm{Earth}=8.5\,\mathrm{kpc}, E=100\,\mathrm{GeV})\sim 5\cdot10^{6}$~yr as the value of the escape time scale as measured at Earth \cite{BeckerTjus2020}. Using $\beta\sim 0.1$ for the inner Galaxy 
and $\gamma_\parallel\sim 0.7$, results in $2-\beta\,\gamma_\parallel =1.93$. This way, we get
\begin{equation}
    \tau_{\rm diff}\sim       10^{7}~\rm{yr}~\left(\frac{r_{\rm gc}}{r_0}\right)^{1.93}\,\left(\frac{E}{10\,\rm GeV}\right)^{-0.7}\,.
\end{equation}
The Galactic wind dominates as a loss process if $\tau_{\rm conv}<\tau_{\rm diff}$, which is the case for energies
\begin{equation}
E\lesssim 700\, {\rm GeV}\, \left(\frac{r_{\rm gc}}{r_0}\right)^{1.93}\,.
\end{equation}
This equation is fulfilled for cosmic-ray energies $E\lesssim 700$~GeV at $r_0$, and still for energies $E\lesssim 10$~GeV in the inner central molecular zone (CMZ), i.e.\ the inner $200$~pc. 
Even by taking the uncertainties of the actual wind speed and the normalization of the diffusion time scale into account, cosmic-ray transport will be influenced by Galactic wind in the CMZ. Therefore, the measured spectral index in the energy range measured with Fermi as shown in Fig.\ \ref{fig:multipanel} can be dominated by the wind as well. Thus, a model in which convection dominates in the inner Galaxy, changing to diffusive escape along the perpendicular direction of the field around $r_{\rm gc}\sim 5$~kpc, fits the data well, with a possible further change to parallel transport in the outermost Galaxy. One prediction of this model is that the transport in the inner Galaxy should change from convection to parallel diffusion for energies larger than $\sim 100$~GeV. This implies a significant change in the energy behavior of the diffuse spectrum from $E^{-2.3}$ to $E^{-3.0}$ at TeV energies. With the emerging Cherenkov Telescope Array (CTA) \cite{cta} and its use of more than a hundred IACTs in the Northern and Southern Hemispheres, the background will be reduced and the field of view increased to $\approx 10$°, while improving sensitivity by an order of magnitude compared to current observatories consisting of only a few ground-based Imaging Air Cherenkov Telescopes (IACTs) \cite{Tibaldo2021}. These enhancements from diffuse measurements of the inner Galaxy with CTA, once available, will allow verification of our model. Complementary high-sensitivity observations of Large High Altitude Air Shower Observatory (LHASSO) \cite{LHAASO2021} and of the Southern Wide-field Gamma-ray Observatory (SWGO) \cite{Albert2019} at higher energies will provide further insights.
\section{Summary and Conclusion}
In this paper, we perform simulations of cosmic-ray diffusion for a range of particle energies and turbulence ratios $b/B$. We analyze the simulation results by only using data that fully lie in the resonant scattering regime, i.e.\ where diffusion can develop fully. In doing so, we find that a power-law fit performs well for the parallel and perpendicular components of the diffusion coefficient, i.e.\ $\kappa_i\propto E^{\gamma_i}$, with $i=\,\parallel,\perp$. We also find that these spectral indices are functions of the turbulence level, $\gamma_i=\gamma_i(b/B)$. In both cases, low turbulence levels lead to the lowest indices, $\gamma_\parallel(b/B=0.07)=0.57$ and $\gamma_\perp(b/B=0.07)=0.33$. Both indices asymptote to $\gamma_i=1$ for values $b/B>1$. This limit is consistent with expectations for $b/B\rightarrow \infty$, as Bohm's theory predicts an index of $1$ for purely turbulent fields. We show that the limit of QLT, which predicts an index for parallel transport of $\gamma_\parallel=1/3$, is not reached at turbulence levels $b/B \gtrsim 0.07$. Since the parallel diffusion coefficients increase significantly as the turbulence level decreases, simulations with even lower turbulence levels require a considerable amount of time for the plateaus of the running diffusion coefficients to develop. Potential numerical effects could then gain influence, which is why much smaller turbulence levels were not considered further \cite{Reichherzer2020}. The gradient of our result indicates that an index of $\gamma_\parallel=1/3$ will be reached at much lower turbulence levels. There, the influence of field-line random walk on the effective perpendicular diffusion process increases. Since the FLRW is energy-independent, $\gamma_\perp$ also becomes smaller for decreasing turbulence levels.

Finally, we apply these results from fundamental plasma physics to the Galaxy. Here, measurements indicate a steepening of the local cosmic-ray energy spectra along the galactocentric radius, with
\begin{equation}
    n\propto \left\{ \begin{array}{cc}
         E^{-2.3}-E^{-2.4}&r_{\rm gc}<5\,\rm kpc  \\
         E^{-2.7}& 5~\rm{kpc}<r_{\rm gc}<10~\rm kpc\\
         E^{-2.9}&r_{\rm gc}\sim 15-20~\rm kpc\\
    \end{array}\right.
\end{equation}
The Kolmogorov approximation in QLT of a diffusion-based change in the energy behavior corresponding to a steepening by $E^{-1/3}$ would certainly be compatible with the inner, flat spectra. However, we were able to show here that for the turbulence level in the inner Galaxy, diffusion would steepen the spectrum significantly more, e.g.\ by $E^{-2/3}$. Our conclusion is therefore that transport in the inner Galaxy must be influenced by a Galactic wind that does not change the energy spectrum, so that the spectrum after transport corresponds to the one at the sources for $r_{\rm gc}<5$~kpc, i.e.\ $n\propto S\propto E^{-2.35\pm0.05}$. The large uncertainties of gamma-ray emission measurements in the inner Galaxy currently provides flexibility in the interpretation of these data, as evidenced by the interpretation of \cite{Recchia2016} of a very soft spectrum and of \cite{Cerri2017} and the current work of a hard proton spectrum in the inner part of the Galaxy. To resolve this tension, better data of the observationally challenging inner Galaxy is needed.

The steepening of the spectral index toward the outer Galaxy would then be due to the change in the transport process, from convective to diffusive. Here, we show that in the next-outer region, diffusive escape is dominated by the perpendicular transport component, with the perpendicular diffusion coefficient steepening the spectrum by $E^{-0.4}$, leading to a spectrum after transport of $n\propto S\,\kappa_{\perp}^{-1}\propto E^{-2.7}$. A dominating perpendicular transport for these galactocentric radii is also suggested by e.g.~\cite{Evoli2012, Cerri2017}. At the outermost radii of the Galaxy, at around $20$~kpc, diffusive escape will again move toward parallel transport and, the spectrum would steepen further. We show in this paper that these findings are consistent with observations at GeV energies. In contrast to \cite{Cerri2017}, we expect a change in the spectrum at around $100$~GeV - $1$~TeV cosmic-ray energies in the inner Galaxy due to the replacement of convective transport by parallel diffusion as the dominant escape process, which should result in a change from $n\propto S\propto E^{-2.35\pm0.05}$ to $n\propto E^{\approx-3}$ according to our simulation results. 
In the future, more detailed simulations, accounting also for adiabatic energy changes that are currently neglected, and improved observations of the diffuse gamma-ray component in the Galaxy with CTA and SWGO over a greater energy range can help to test this picture and will help to discriminate the three discussed models.



\section*{Declarations}
\subsection*{Funding}
This work is supported by the ``ADI 2019’’ project funded by the IDEX Paris-Saclay, ANR-11-IDEX-0003-02 (PR). PR also acknowledges support by the German Academic Exchange Service and by the RUB Research School via the \textit{Project International} funding. LM acknowledges financial support from the Austrian Science Fund (FWF) under grant agreement number I 4144-N27. EZ gratefully acknowledges support by NSF AST2007323.
 
%
\subsection*{Conflict of interest}
On behalf of all authors, the corresponding author states that there is no conflict of interest. 

\subsection*{Availability of data}
The methods, results, and data in this paper are fully available to interested researchers.

\subsection*{Code availability}
Simulations were performed with the publicly available tool CRPropa \cite{AlvesBatista2016} (the specific version used for the simulations is CRPropa 3.1-f6f818d36a64), supported by various analysis tools: NumPy~\cite{van_der_Walt_2011}, Matplotlib~\cite{Hunter_2007}, Pandas~\cite{pandas}, and jupyter notebooks~\cite{jupyter-notebook}.

\bibliographystyle{spphys}       

\end{document}